Classification: Physical Sciences: Physics

# Continuous and Discontinuous Quantum Phase Transitions in a Model Two-Dimensional Magnet

S. Haravifard[a,b], A. Banerjee[b], J. C. Lang[a], G. Srajer[a], D. M. Silevitch[b], B. D. Gaulin[c,d], H. A. Dabkowska[d], and T. F. Rosenbaum[b]

[a] Advanced Photon Source
 Argonne National Laboratory
 9700 S. Cass Avenue
 Argonne, Illinois 60439, USA
[b] James Franck Institute and Department of Physics
 The University of Chicago
 929 E 57th Street
Chicago, Illinois 60637, USA
[c] Department of Physics and Astronomy
 McMaster University
1280 Main St. W, Hamilton
 Hamilton, Ontario, L8S 4M1, Canada
and
Canadian Institute for Advanced Research
180 Dundas St. W.
Toronto, Ontario, M5G 1Z8, Canada
[d] Brockhouse Institute for Material Research
 McMaster University
1280 Main St. W, Hamilton
 Hamilton, Ontario, L8S 4M1, Canada

Corresponding author: T.F. Rosenbaum
The University of Chicago
929 E. 57th Street
Chicago, IL 60637
773-702-7256
tfr@uchicago.edu

16 pages, 4 figures

The Shasty-Sutherland model, which consists of a set of spin 1/2 dimers on a 2-dimensional square lattice, is simple and soluble, but captures a central theme of condensed matter physics by sitting precariously on the quantum edge between isolated, gapped excitations and collective, ordered ground states. We compress the model Shastry-Sutherland material, $SrCu_2(BO_3)_2$, in a diamond anvil cell at cryogenic temperatures to continuously tune the coupling energies and induce changes in state. High-resolution x-ray measurements exploit what emerges as a remarkably strong spin-lattice coupling to both monitor the magnetic behavior and the absence or presence of structural discontinuities. In the low-pressure spin-singlet regime, the onset of magnetism results in an expansion of the lattice with decreasing temperature, which permits a determination of the pressure dependent energy gap and the almost isotropic spin-lattice coupling energies. The singlet-triplet gap energy is suppressed continuously with increasing pressure, vanishing completely by 2 GPa. This continuous quantum phase transition is followed by a structural distortion at higher pressure.

**Introduction**

One of the first real-world examples encountered in elementary quantum mechanics is the hydrogen atom, where two correlated spin 1/2 particles, the electron and the proton, combine in multiple configurations to define a set of discrete energy levels. The ground state singlet superposes up and down configurations of the spins to yield a state with total spin zero; the excited state triplet provides three ways to produce a total spin of one. The fundamental quantum mechanics of spin singlets physically placed on ladders or constrained to sheets is a natural extension, with a rich set of possibilities for collective, interacting states from spin liquids to magnets to exotic superconductors (1-4). A particularly important topology involves spin 1/2 orthogonal dimers on a square lattice, the so-called Shastry-Sutherland model (5), because it is exactly soluble. Here we exploit a remarkably strong coupling of the spin degrees of freedom to the lattice in a model Shastry-Sutherland system to probe the evolution of the collective spin configurations with temperature and pressure using x-rays and diamond anvil cell technology. Thermal excitations from singlet to triplet define a gap, which disappears as pressure P tunes the ratio between inter- and intra-dimer coupling energies. This continuous quantum phase transition occurs at P ~ 2 GPa, followed by a first order transition at P ~ 4.5 GPa, which we associate with the full-fledged onset of long-range antiferromagnetic order.

Insulating $SrCu_2(BO_3)_2$ (SCBO) successfully captures the physics of the Shastry-Sutherland model(6), with corner-sharing $Cu^{2+}$ S=1/2 dimers lying on a square lattice and only weak interactions between adjacent planes. Under ambient conditions, the system has tetragonal symmetry with $a = 8.995$ Å and $c = 6.649$ Å. ESR (7) and neutron (8) data establish that the spin interactions in SCBO are well described by the Shastry-Sutherland Hamiltonian:

$$H = J \sum_{nn} S_i \cdot S_j + J' \sum_{nnn} S_i \cdot S_j \qquad , \qquad (1)$$

where $J$ and $J'$ are the antiferromagnetic intra- and inter-dimer exchange interactions, respectively. $J$ has been measured to lie between 84 and 100 K, with $x \equiv J'/J$ between 0.63 and 0.68 (9-11), just below the predicted critical value of $x = 0.69$ for the quantum phase transition (12). Initially, it was posited that this T = 0 transition with $x$ would take the system from the gapped singlet state directly into the Néel state of long-range antiferromagnetic order (10). However more recent theoretical work has suggested that the transition at $x = 0.69$ brings the system into an intermediate state composed of local plaquettes of ordered spins, followed by a second transition at higher $x$ into the Néel state (12, 13).

Prior experiments have investigated the use of Al, La, Na and Y substitution for Sr as a potential source of chemical pressure (14), along with Mg substituting for Cu as a source of magnetic dilution (15), to drive SCBO across the $T = 0$ phase boundary. Chemical doping, however, necessarily introduces a significant degree of disorder into the system, disrupting the formation of long-range order like the predicted four sublattice Néel state and possibly altering the quantum critical response. The need for only a slight change in $x$ to drive the system across the quantum phase transition, combined with the exponential dependence of the dimer exchange couplings of Eq. (1) on separation, points to pressure as an alternative tuning parameter. Hydrostatic pressure preserves the underlying Hamiltonian, avoiding the introduction of disorder or any symmetry-breaking fields, and it permits the study of the same physical sample across a range of values of $x$. Previous magnetic susceptibility measurements of pure SCBO under pressure (16), while not reaching the quantum critical point, offer an estimate of the required pressure scale to be approximately 2 to 3 GPa, with room temperature results on powdered samples pointing to a structural transition at significantly higher pressures (17).

Applying truly hydrostatic pressures up to 6 or 8 GPa at cryogenic temperatures can be achieved with the use of a diamond anvil cell (18), but with limits on sample size to sub-millimeter dimensions and prohibitive attenuation of such natural magnetic probes as neutrons. Instead, we turn to high-energy synchrotron x-ray scattering (see Methods), and we approach the question of the magnetic response via precise measurements of the crystal lattice parameters. This method presumes a strong coupling between the spin states and the lattice, seen for example in low-dimensional quantum spin systems such as quasi-1D spin-Peierls systems (19). Indeed, recent magnetostriction measurements of SCBO in pulsed fields (20) point to a sizeable coupling between the spins in the Cu dimers and the lattice. Powder neutron measurements (21) observe structural changes such as shifts in the Cu-O-Cu intraplane bond angle coincident with the onset of the single-triplet state. Moreover, inelastic neutron scattering (22), infrared spectroscopy with polarized light (23), and comparisons of the elastic constants with the magnetic susceptibility (24, 25) all support the proposition that a close study of the lattice can offer powerful insights into the magnetic state of the system.

**Results and Discussion**

We explicitly demonstrate in Fig. 1 that the variation of the lattice with $P$ and $T$ reveals the essential aspects of the low temperature spin response. The inverse of the lattice constant $a$, measured using the (400) Bragg peak, scales directly with the magnetic susceptibility results reported by Waki *et al.* (16). Both measurements reveal the transition with $T$ from the collective ground state singlet to the excited state triplet. Our lattice constant data is multiplied by $5.15 \times 10^{-6}$ to match the susceptibility curves at each coincident $P$, corresponding to a pressure-independent spin-lattice coupling constant $a(d\Delta/da) \sim 10^5$ K, where $\Delta(P)$ is the energy gap between the singlet and triplet states. The lattice actually expands with decreasing temperature because of the magnetic transition from triplet to singlet. The strong spin-lattice coupling in SCBO provides a powerful tool to use precision measurements of the lattice constants to discern the underlying nature of the magnetic state and, as discussed below, to distinguish between theoretical predictions involving short-range and long-range magnetic order. .

We plot in Fig. 2 the temperature variation of the planar lattice constant for pressures extending up to 2.8 GPa. In order to remove the purely structural compression of the lattice with increasing $P$, thereby isolating the magnetic contribution, the data in Fig. 2 is normalized by the compression at $T$ = 20 K: $\omega \equiv a(P = 0, T = 20K)/a(P, T = 20K)$. Neutron measurements at ambient P show that by $T$ = 20 K all the dimer singlets have melted and the system is solely in the triplet state (22); it is also a temperature sufficiently cold that thermal expansion between 20 K and base is negligible. The resulting curves trace out the singlet-triplet gap, which closes with increasing $P$. Above 2 GPa, the gapped behavior vanishes, signifying that the system has been driven from the gapped singlet state into a different ground state. The cartoons of Fig. 3A illustrate the predicted evolution with $P$ of the ground state from collective non-magnetic singlet to local magnetic plaquettes to full-blown antiferromagnetic order in the form of four interpenetrating Néel lattices(12, 13). The intermediate state is predicted to consist of 4-site plaquettes, each of which acts as an effective singlet(12).

The low-temperature magnetic response of SCBO over the full $P$-$T$ plane is captured in the color plot of Fig. 3B. We show a normalized triplet density obtained from the scaled inverse lattice constant $\frac{1}{\omega a}$, using the normalization described above to isolate the magnetic response from the background pressure-dependent mechanical compression, followed by a linear scaling to values between 0 and 1. At low pressures, the presence of the gap is clearly visible, along with the continuous suppression of the gapped state as $P$ increases. At a critical pressure, the gapped behavior vanishes and the system enters a new state where the magnetism is largely temperature-independent over the range studied here. We superpose in Fig. 3B values for the gap deduced from the temperature dependence of the inverse lattice constant (Fig. 1), adopting the same thermally-activated form, $\alpha + \beta \exp(-\Delta/kT)$, used to describe the magnetic susceptibility results. At ambient pressure, we find a gap $\Delta$ = 2.45±0.08 meV, consistent with magnetic susceptibility measurements which found $\Delta$ = 2.41±0.01 meV (26) and somewhat smaller than the 2.8±0.1 meV measured via inelastic neutron scattering (22). The gap closes linearly with increasing $P$ down to an energy of 0.36 meV at 1.75 GPa. Extrapolating the linear trend to zero yields an estimate for the critical pressure of $P_c$ = 1.93± 0.07 GPa. This value for the critical pressure is in accord with lower resolution results derived from neutron scattering experiments (27), although there the inferred values of the gap only show it decreasing to 1.7 meV at $P$ = 1.9 GPa.

An alternative estimate for $P_c$ can be obtained by taking a base temperature cut through the color map of Fig. 3B, defining $P_c$ as the point where the triplet density goes to zero. Here, we find $P_c$ = 2.05 ± 0.1 GPa. These values are consistent within error bars, but sub-Kelvin temperature measurements would be required to establish firmly a critical exponent of 1 for the $T$ = 0 pressure variation of the gap. $P_c$ defines the low-temperature limit of the phase boundary. It is reasonable to assume that pressures above $P_c$ will stabilize the second phase at finite temperature. Since the singlet-triplet gap has been suppressed to zero, and in the absence of a long-range ordered magnetic state that can couple to the lattice, our measurements are unable to resolve the shape or scale of this phase boundary.

As noted above, we observe an expansion of the lattice with decreasing $T$ through the triplet-singlet transition in the $ab$ plane. When normalized by the lattice constant to get a dimensionless magnetostriction, $\varepsilon$, the per-unit-cell coupling strength $(d\Delta/d\varepsilon)$ is ~ 1.5×$10^5$ K, where $\Delta$ = 2.45 meV is the ambient-pressure spin gap. This coupling energy is comparable in magnitude to what has been observed in many oxide magnets, for instance the spin-Peierls material $CuGeO_3$ (28). The observed striction in the $ab$ plane can be compared with striction measurements as a function of applied magnetic field (29), where complete excitation into the triplet state resulted in qualitatively similar behavior, but only ¼ the magnitude, pointing to a strong field dependence of the spin-lattice coupling.

Additionally, we can derive an estimate for the energy contained in the magnetic excitations by comparison to the compression of the *ab* plane from the application of pressure. The contraction of the lattice due to the triplet excitation corresponds to an effective pressure $P_{mag} = \lambda(\Delta A_{mag}/A) \sim 0.2$ GPa, where $\Delta A = (A_{20K} - A_{base})|_{P=0,}$ and the section modulus $\lambda$ was measured at $T = 20$ K to be 450 GPa; the 20 K bulk modulus was measured to be 76 GPa. This effective pressure gives a magnetic energy density of 0.150 meV per unit cell, taking $c$=6.643 Å. The continuous nature of the quantum phase transition at $P_c = 2.0$ GPa can be quantified from the measurements of *a(P,T~4K)* for $0 < P < 6.2$ GPa in Fig. 4A. There is no discernible break in *a(P)* at $P_c$ and separate linear best fits for $0 < P < 2$ GPa and $2 < P < 4$ GPa bound the change in the bulk modulus to be less than 24% with 90% confidence at the quantum phase transition. The continuity of both the value and the slope of the lattice constant across the phase boundary provides insight into the nature of the intermediate magnetic state. If the magnetic state exhibited long range order, then a strain field would be applied to the lattice via the strong observed spin-lattice coupling in SCBO, giving rise to an effective $\phi^4$ term in the free energy of the lattice. The result would be a change in the bulk modulus of the lattice at the transition. Similarly, if the magnetic transition were first order, the same coupling would be expected to produce a discontinuity in the pressure dependence of the lattice constant. Experimentally, neither the lattice constant nor the bulk modulus exhibits a discontinuity at the critical pressure. The transition is second order, and we conclude that the intermediate magnetic state is characterized by short-range magnetic order. This conclusion is consistent with the theoretical predictions for the intermediate state of a set of plaquettes with no effective inter-plaquette ordering(12). It is in principle possible to estimate the characteristic length scale of this intermediate state by tracking the amount of broadening of the lattice Bragg peaks, however for these measurements, history-dependent strains induced by small deviations from true hydrostatic pressure dominate the linewidths.

We contrast the results with increasing pressure at low temperature to the magnetic states observed in SCBO when high magnetic fields are used to close the singlet-triplet gap (30). The NMR measurements reported in Ref. (30) show a magnetic state comprised of a complex spin superlattice of triplets with an extended unit cell, consistent with theoretical models for the magnetized state (31, 32). Importantly, the transition into this ordered state was observed to be first order. Moreover, the x-ray measurements of Ref. (29) show that the lattice continues to contract with increasing magnetic field above the 20 T transition – the signature of the long-range ordered superlattice – while our x-ray measurements demonstrate that, to the contrary, the lattice constant no longer evolves with increasing pressure between 2 and 4.5 GPa – the mark of the short-range ordered plaquette state. While our x-ray measurements cannot resolve the detailed microstructure of this intermediate state, it is clear that tuning the lattice constants of SCBO with pressure accesses a different set of states than application of a magnetic field.

At higher pressures (and hence higher $x$), SCBO is predicted to undergo a second quantum phase transition out of the plaquette state and into the Néel state (Fig. 3A). Room temperature powder diffraction results reported by Loa et al. (17) show a second-order tetragonal to monoclinic transition at $P = 4.7$ GPa. This monoclinic transition was subsequently reported to be weakly temperature dependent, varying between 4.2 and 4.8 GPa for temperatures ranging from room temperature down to 30 K (27), although the order of the transition was not reported. Our results at low $T$ contrast sharply with the behavior reported at room temperature. The single-crystal measurements of $a$ at $T \sim 4$ K reveal a first-order structural transition at approximately the same

*P*, marked by a discontinuous decrease in the lattice constant of 0.5%. The same amplitude discontinuity is observed at $T \sim 20$ K. With the strong spin-lattice coupling in SCBO it is to be expected that a reconfiguration of the magnetic state, particularly with the emergence of long-range order, will be linked to the crystal structure. Finding a definitive signature of antiferromagnetism at half-integral reciprocal lattice positions is the subject of ongoing investigations.

The pressure dependence of the gap can be used to estimate the fundamental scaling parameter from theory, *Δx(P)*, by using the dimer expansion model described in (33) under the simplifying assumption that *J* is approximately constant with pressure. As shown in Fig 4B, *Δx* varies approximately exponentially with pressure in the dimer phase, changing the requisite amount to cross the quantum critical point. If the exponential dependence holds to higher *P*, it would imply a much smaller variation between the continuous and discontinuous quantum transitions. A more complete calculation of *x* would require an independent measurement of the pressure evolution of either *J* or *J'*, and would also need to incorporate the *J''* inter-plane interactions (33), in particular the Dzyaloshinsky-Moria interactions between adjacent planes (34).

The Shastry-Sutherland model and its material manifestation SCBO serve as a physical palimpsest for the complexities of correlated systems with competing ground states. A deceptively simple model of coupled spin ½ singlets arranged on a square lattice reveals the subtle interaction of structure and quantum magnetism, with multiple T = 0 transitions that balance local and long-range magnetic order with preserved and broken lattice symmetries. The apparent formation of plaquettes before the full Néel state emerges is reminiscent of the tendency of reduced dimensionality quantum systems towards microphase separation, with general implications for metal-insulator transitions and superconductivity (35). From an experimental perspective, the salubrious match of high-resolution x-ray probes and diamond anvil cell technology with systems prone to strong spin-lattice coupling opens up new approaches to cleanly tuning magnetic quantum phase transitions with high-quality single crystals of $<10^{-6}$ cm$^3$ volumes.

**Methods**

Single crystals of SCBO were grown using a floating-zone image furnace method (36) and cleaved to approximately (50x50x30) μm$^3$. Individual samples were loaded into a diamond anvil cell (DAC) using a 4:1 Methanol:Ethanol pressure medium, which provides hydrostatic pressure in the range of interest at cryogenic temperatures. The DAC was fitted with a helium membrane for in-situ pressure tuning; pressure was measured via a lattice measurement of a piece of polycrystalline silver placed close to the sample inside the cell (18). The high pressure measurements were performed at Sector 4-ID-D of the Advanced Photon Source (APS) at Argonne National Laboratory (ANL) using a 6-circle diffraction stage, while the ambient-pressure measurements were performed at Sector 4-ID-B of APS/ANL. The data presented here were obtained from 6 samples over the course of 3 beam runs. The lattice constant $a$ was measured using the (400) Bragg peak; $c$ was measured using a combination of the (301), (042), and (52±1) peaks. Due to the location of the thermometer on the cryostat, the data has been shifted upwards in temperature by 1.9 K to correspond with the accepted ambient-pressure temperature scale of SCBO.

**Acknowledgments**  We are grateful to Y. Feng for helpful discussions. The work at the University of Chicago was supported by National Science Foundation (NSF) Grant DMR-0907025. D.M.S. acknowledges support from US Department of Energy (DOE), Basic Energy Sciences Grant DEFG02-99ER45789. Use of APS is supported by the US DOE Office of Basic Energy Sciences.

## References

1. Anderson P (1987) The Resonating Valence Bond State in $La_2CuO_4$ and Superconductivity. *Science* 235:1196–1198.

2. Dagotto E, Rice T (1996) Surprises on the way from one- to two-dimensional quantum magnets: The ladder materials. *Science* 271:618–623.

3. Lee SH et al. (2007) Quantum-spin-liquid states in the two-dimensional kagome antiferromagnets $Zn_xCu_{4-x}(OD)_6Cl_2$. *Nature Materials* 6:853–857.

4. Uji S et al. (2001) Magnetic-field-induced superconductivity in a two-dimensional organic conductor. *Nature* 410:908–910.

5. Shastry BS, Sutherland B (1981) Exact ground state of a quantum mechanical antiferromagnet. *Physica B* 108:1069–1070.

6. Kageyama H, Yoshimura K, Kato M, Kosuge K (1999) Exact Dimer Ground State and Quantized Magnetization Plateaus in the Two-Dimensional Spin System $SrCu_2(BO_3)_2$. *Phys Rev Lett* 82:3168–3171.

7. Nojiri H, Kageyama H, Onizuka K, Ueda Y, Motokawa M (1999) Direct Observation of the Multiple Spin Gap Excitations in Two-Dimensional Dimer System $SrCu_2(BO_3)_2$. *J Phys Soc Jpn* 68:2906–2909.

8. Kageyama H et al. (2000) Direct Evidence for the Localized Single-Triplet Excitations and the Dispersive Multitriplet Excitations in $SrCu_2(BO_3)_2$. *Phys Rev Lett* 84:5876–5879.

9. Zheng W, Hamer C, Oitmaa J (1999) Series expansions for a Heisenberg antiferromagnetic model for $SrCu_2(BO_3)_2$. *Phys Rev B* 60:6608–6616.

10. Miyahara S, Ueda K (1999) Exact dimer ground state of the two dimensional Heisenberg spin system $SrCu_2(BO_3)_2$. *Phys Rev Lett* 82:3701–3704.

11. Totsuka K, Miyahara S, Ueda K (2001) Low-Lying Magnetic Excitation of the Shastry-Sutherland Model. *Phys Rev Lett* 86:520–523.

12. Koga A, Kawakami N (2000) Quantum Phase Transitions in the Shastry-Sutherland Model for $SrCu_2(BO_3)_2$. *Phys Rev Lett* 84:4461–4464.

13. Miyahara S, Ueda K (2003) Theory of the orthogonal dimer Heisenberg spin model for $SrCu_2 (BO_3)_2$. *Journal of Physics: Condensed Matter* 15:R327–R366.

14. Liu G et al. (2005) Doping effects on the two-dimensional spin dimer compound $SrCu_2(BO_3)_2$. *Phys Rev B* 71:014441.

15. Haravifard S et al. (2006) In-Gap Spin Excitations and Finite Triplet Lifetimes in the

Dilute Singlet Ground State System $SrCu_{2-x}Mg_x(BO_3)_2$. *Phys Rev Lett* 97:247206.

16.  Waki T et al. (2007) A Novel Ordered Phase in $SrCu_2(BO_3)_2$ under High Pressure. *Journal of the Physical Society of Japan* 76:073710.

17.  Loa I et al. (2005) Crystal structure and lattice dynamics of $SrCu_2(BO_3)_2$ at high pressures. *Physica B: Condensed Matter* 359-361:980–982.

18.  Feng Y, Jaramillo R, Wang J, Ren Y, Rosenbaum TF (2010) Invited Article: High-pressure techniques for condensed matter physics at low temperature. *Rev. Sci. Instrum.* 81:041301.

19.  Lumsden M, Gaulin B (1999) Critical phenomena at the spin-Peierls transition in $MEM(TCNQ)_2$. *Phys Rev B* 59:9372–9381.

20.  Jorge GA et al. (2005) Crystal symmetry and high-magnetic-field specific heat of $SrCu_2(BO_3)_2$. *Phys Rev B* 71:092403.

21.  Vecchini C et al. (2009) Structural distortions in the spin-gap regime of the quantum antiferromagnet $SrCu_2(BO_3)_2$. *J Solid State Chem* 182:3275–3281.

22.  Gaulin B et al. (2004) High-Resolution Study of Spin Excitations in the Singlet Ground State of $SrCu_2(BO_3)_2$. *Phys Rev Lett* 93:267202.

23.  Homes C et al. (2009) Infrared spectra of the low-dimensional quantum magnet $SrCu_2(BO_3)_2$: Measurements and ab initio calculations. *Phys Rev B* 79:125101.

24.  Zherlitsyn S et al. (2000) Sound-wave anomalies in $SrCu_2(BO_3)_2$. *Phys Rev B* 62:R6097–R6099.

25.  Wolf B et al. (2001) Soft Acoustic Modes in the Two-Dimensional Spin System $SrCu_2(BO_3)_2$. *Phys Rev Lett* 86:4847–4850.

26.  Aczel A et al. (2007) Impurity-induced singlet breaking in $SrCu_2(BO_3)_2$. *Phys Rev B* 76:214427.

27.  Zayed M (2010) *Novel States in Magnetic Materials under Extreme Conditions: A High Pressure Neutron Scattering Study of the Shastry-Sutherland compound $SrCu_2(BO_3)_2$* (ETH Zurich).

28.  Boucher JP, Regnault LP (1996) The Inorganic Spin-Peierls Compound $CuGeO_3$. *J. Phys. I France* 6:1939–1966.

29.  Narumi Y et al. (2009) Field Induced Lattice Deformation in the Quantum Antiferromagnet $SrCu_2(BO_3)_2$. *J Phys Soc Jpn* 78:043702.

30.  Kodama K et al. (2002) Magnetic superstructure in the two-dimensional quantum

antiferromagnet SrCu$_2$(BO$_3$)$_2$. *Science* 298:395–399.

31. Momoi T, Totsuka K (2000) Magnetization plateaus as insulator-superfluid transitions in quantum spin systems. *Phys Rev B* 61:3231–3234.

32. Miyahara S, Ueda K (2000) Superstructures at magnetization plateaus in SrCu$_2$(BO$_3$)$_2$. *Phys Rev B* 61:3417–3424.

33. Zheng W, Oitmaa J, Hamer C (2001) Phase diagram of the Shastry-Sutherland antiferromagnet. *Phys Rev B* 65:014408.

34. Cépas O et al. (2001) Dzyaloshinski-Moriya Interaction in the 2D Spin Gap System SrCu$_2$(BO$_3$)$_2$. *Phys Rev Lett* 87:167205.

35. Dagotto E (2005) Complexity in strongly correlated electronic systems. *Science* 309:257–262.

36. Dabkowska HA et al. (2007) Crystal growth and magnetic behaviour of pure and doped SrCu$_2$($^{11}$BO$_3$)$_2$. *J Cryst Growth* 306:123–128.

**Figure Legends**

**Fig 1** Parallel evolution with temperature $T$ of x-ray lattice constant measurements and magnetic susceptibility measurements for $SrCu_2(BO_3)_2$ under pressure. Left axis: inverse lattice constant $1/a$, determined from the (400) Bragg reflection. 0.9 and 1.4 GPa traces have been vertically offset by $5.86 \times 10^{-5}$ and $1.21 \times 10^{-4}$ Å$^{-1}$ respectively to account for the overall compression of the sample as pressure is applied. Right axis: Magnetic susceptibility, from Waki et al.(16).

**Fig 2** Lattice constant $a$ of $SrCu_2(BO_3)_2$ as a function of temperature for a range of pressures. Traces have been normalized using the 20 K compressibility curve, as described in the text. The presence of the gapped-singlet state is visible at low pressure, and is suppressed by 2 GPa, driving the system into a different ground state.

**Fig 3** Pressure-driven magnetic phase diagram for $SrCu_2(BO_3)_2$. **A** Schematic showing the predicted ground states of $SrCu_2(BO_3)_2$ in three pressure regimes. Left to right, the system is in a singlet state with a gap to a triplet, a plaquette-based state, and an antiferromagnetic state consisting of four interpenetrating sublattices. Solid lines represent the intradimer ($J$) couplings; dashed lines the interdimer ($J'$) couplings. Arrows represent local effective singlets formed by each plaquette. **B** Normalized triplet density as a function of temperature and pressure, using the normalization described in the text. At low temperature, the triplet density is suppressed with increasing pressure, going to zero at a $P_c$ comparable to the pressure where the gap extrapolates to zero. Filled circles show calculated values of the low temperature singlet gap (right axis), based on a thermally activated model.

**Fig 4** High pressure behavior of $SrCu_2(BO_3)_2$. **A** Lattice constant evolution up to high pressure, showing continuous behavior through the region where the spin gap closes, followed by a structural phase transition at 4.5-4.7 GPa, associated theoretically with a transition into a fully ordered antiferromagnetic state. Room temperature powder data from I. Loa et al. (17) does not pick up the pronounced change in lattice constant. Low temperature data varies by ±0.5 K from pressure to pressure. **B** Evolution of the fundamental coupling ratio $\Delta x / x \equiv (x(P) - x(P=0)) / x(P=0)$, determined from the gap values in the dimer phase. Line is a phenomenological exponential fit to the measured values, with the dashed line showing an extrapolation to the second transition at $P$ = 4.5 GPa. Arrows indicate the pressure values at which the two phase transitions occur.

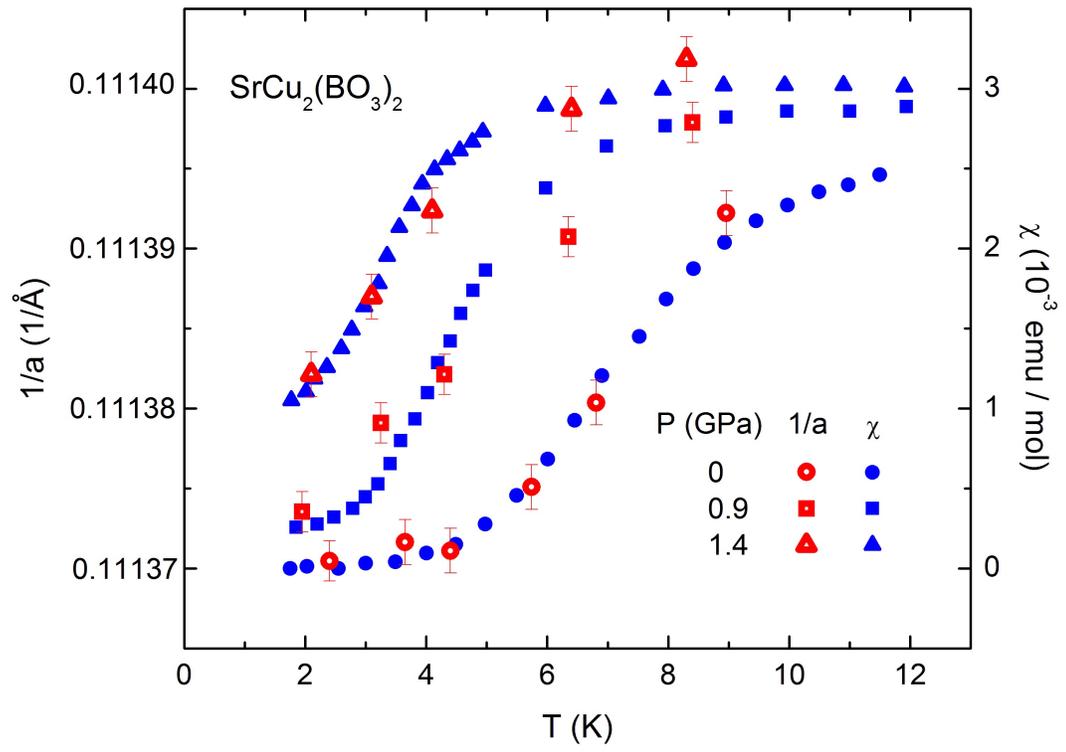

Fig. 1

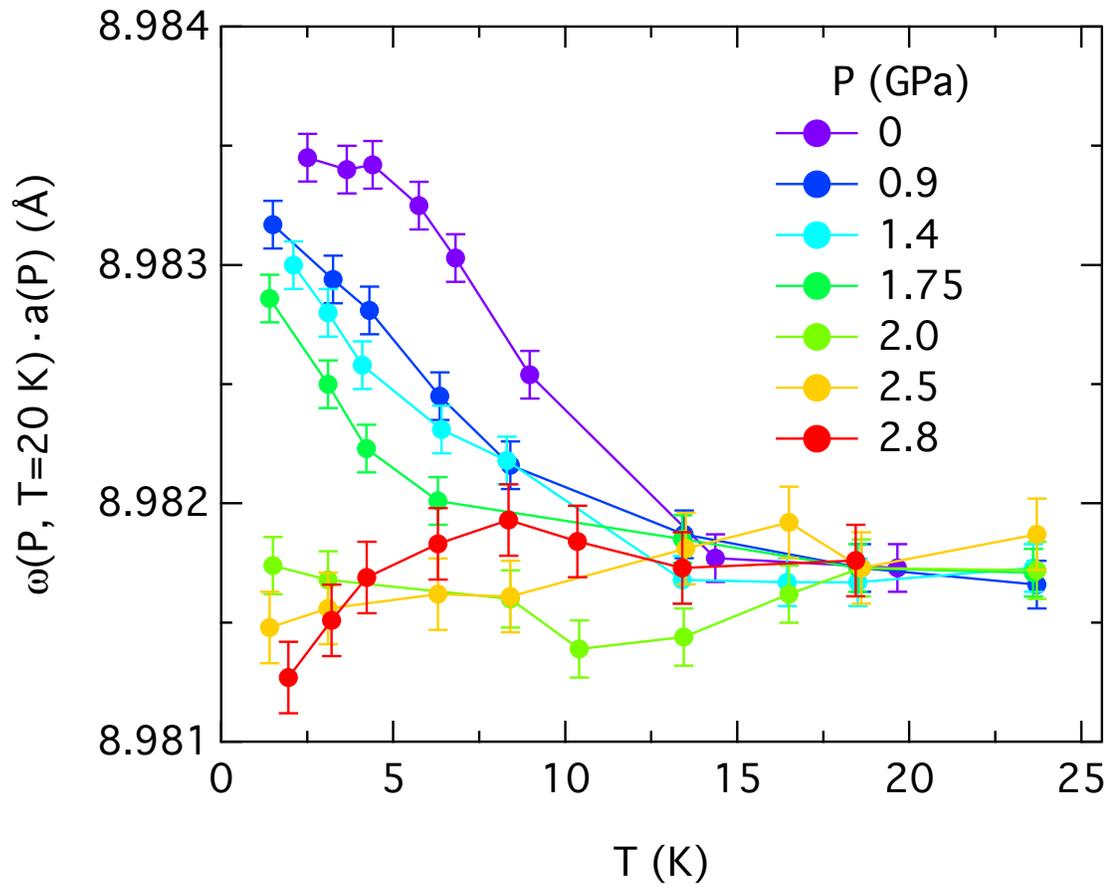

Fig. 2

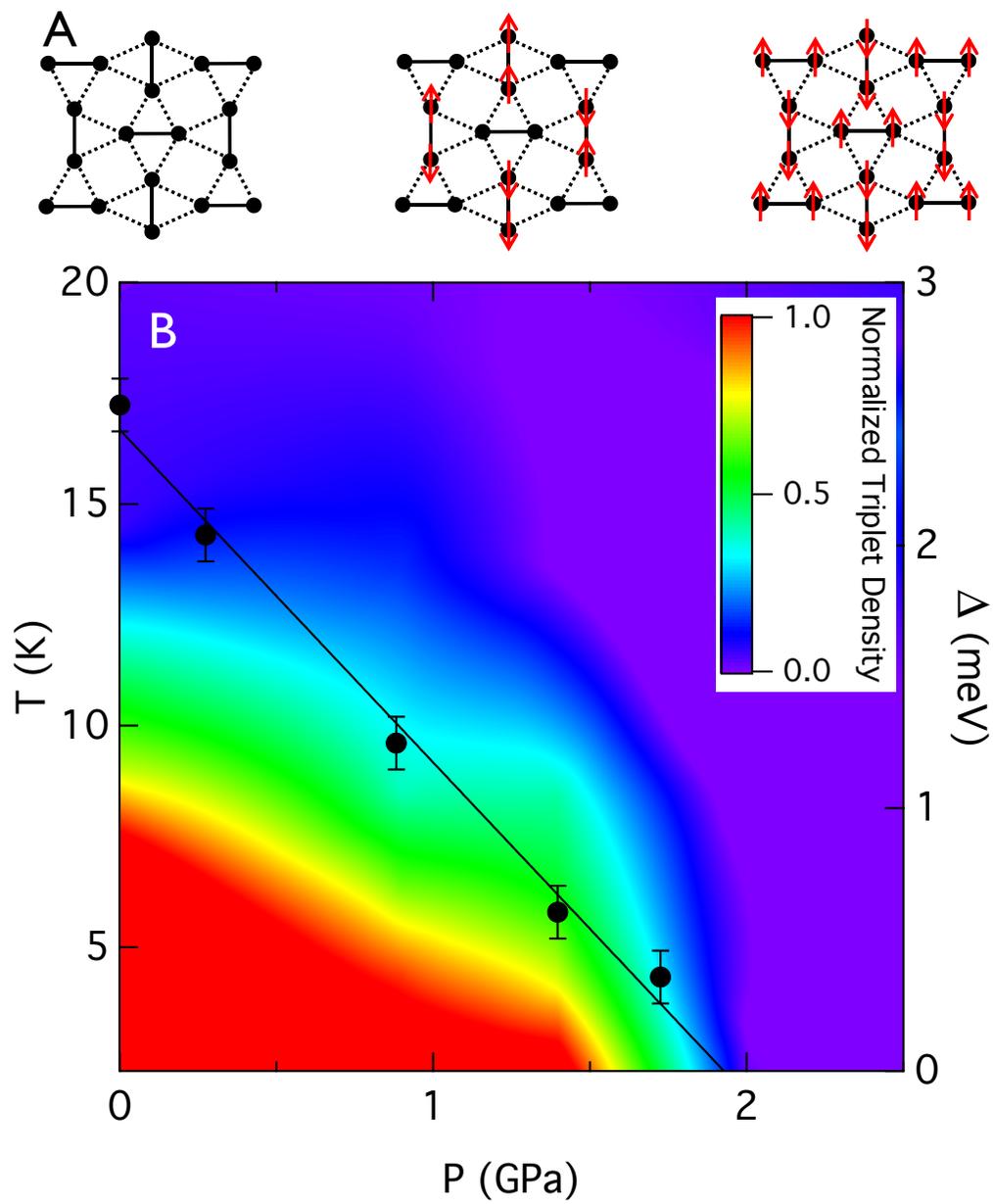

Fig 3

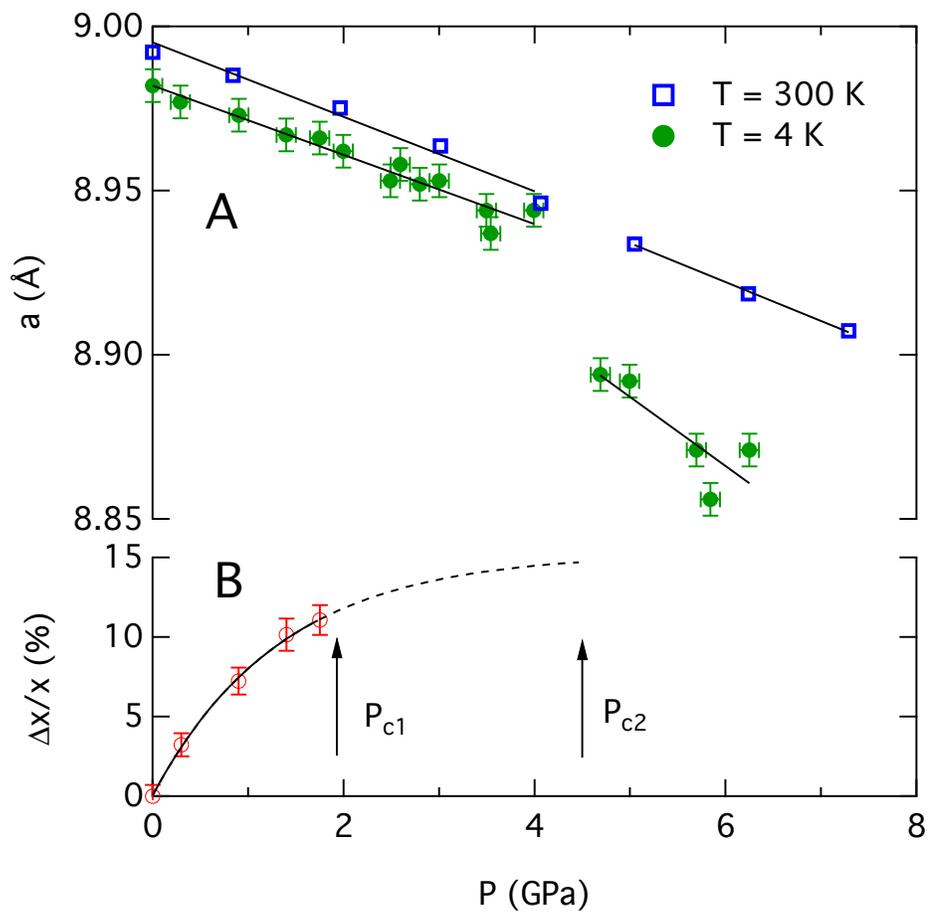

Fig. 4